\documentclass[10pt
]{article}

\usepackage{amsmath}
\usepackage{amssymb}

\usepackage{graphicx}

\usepackage{cite}

\usepackage[labelfont=bf,labelsep=period,justification=raggedright]{caption}

\makeatletter
\renewcommand{\@biblabel}[1]{\quad#1.}
\makeatother

\date{}

\usepackage[]{hyperref}
\hypersetup{colorlinks=true}
\usepackage{color}

\begin{document}

\begin{flushleft}
{\Large
\textbf{Growth patterns and scaling laws governing AIDS epidemic in Brazilian cities}}
\\
F. J. Antonio\textsuperscript{1\textasteriskcentered},
S. Picoli\textsuperscript{1},
J. J. V. Teixeira\textsuperscript{2},
R. S. Mendes\textsuperscript{1}
\\
\bf{1}~Departamento de F\'{\i}sica, Universidade Estadual de Maring\'{a} \\ \textit{and} National Institute of Science and Technology for Complex Systems, Avenida Colombo 5790, 87020-900, Maring\'{a}~--~PR, Brazil
\\
\bf{2}~Departamento de An\'{a}lises Cl\'{\i}nicas e Biomedicina, Universidade Estadual de Maring\'{a}, Avenida Colombo 5790, 87020-900, Maring\'{a}~--~PR, Brazil
\\
\textasteriskcentered~E-mail:~fja@dfi.uem.br
\end{flushleft}

\section*{Abstract}

Brazil holds approximately 1/3 of population living infected with AIDS (acquired immunodeficiency syndrome) in Central and South Americas, and it was also the first developing country to implement a large-scale control and intervention program against AIDS epidemic. 
In this scenario, we investigate the temporal evolution and current status of the AIDS epidemic in Brazil. 
Specifically, we analyze records of annual absolute frequency of cases for more than 5000 cities for the first 33 years of the infection in Brazil. 
We found that
\textit{(i)} the annual absolute frequencies exhibit a logistic-type growth with an exponential regime in the first few years of the AIDS spreading; 
\textit{(ii)} the actual reproduction number decaying as a power law; 
\textit{(iii)} the distribution of the annual absolute frequencies among cities decays with a power law behavior; 
\textit{(iv)} the annual absolute frequencies and the number of inhabitants have an allometric relationship; 
\textit{(v)} the temporal evolution of the annual absolute frequencies have different profile depending on the average annual absolute frequencies in the cities. 
These findings yield a general quantitative description of the AIDS infection dynamics in Brazil since the beginning. 
They also provide clues about the effectiveness of treatment and control programs against the infection, that has had a different impact depending on the number of inhabitants of cities. 
In this framework, our results give insights into the overall dynamics of AIDS epidemic, which may contribute to select empirically accurate models.

\section*{Introduction}

The spreading of epidemics is a threat to the world public health. 
It has also been arising large scientific attention due to the potential harm to the society~\cite{kreiss1986,morris1997,newman2002,riley2003,lipsitch2003,neumann2009}. 
In particular, the infection by the human immunodeficiency virus (HIV)~---~which causes AIDS~---~has been the subject of intense studies since the early 1980s, when the virus started spreading quickly throughout the globe and has become a worldwide concern~\cite{curran1988,bor1993,vogl1999,weiss2006,fonseca2007,walker2008,hall2010}. 
Estimates reveal that by the end of 2010 more than 30 million people were living with the infection worldwide. 
As a comparison, about 2.7 million new infections were identified on the globe just in 2010~\cite{UNAIDS,who2011}.

The study of the epidemiology of HIV infection using accurate surveillance data has been providing essential information in guiding rational control and intervention programs, as well as in monitoring trends of the epidemic in many countries~\cite{curran1988,amundsen2004,weiss2006,walker2008,beyrer2010,hall2010,alsallaq2013,UNAIDS}. 
For example, it has been shown that new HIV infections among adults depend on their geographic region in the globe and social contact groups distributions~\cite{quinn1996}. 
Brazil~---~the world's fifth most populous country, with almost 200 million inhabitants~---~was the first developing country to implement a large-scale control and intervention programs~\cite{meiners2011,who2011}. 
The Brazilian response to AIDS has included effective prevention strategies with strong participation of the civil society, multi-sectoral mobilization, and intense use of antiretroviral therapies~\cite{bailey1988, fonseca2007}. 
In a global scenario, these facts has attracted significant attention from the scientific community to the dynamics of HIV/AIDS epidemic in Brazil~\cite{teixeira2004,UNAIDS}.

In the present work, we investigate the evolution and current status of the AIDS epidemic in Brazil by analyzing its annual absolute frequency per city. 
Our analysis points in the direction of an alternative approach to the usual methods of epidemics analysis, emphasizing how the dynamics of the virus at the level of municipalities behaves collectively. 
The data were obtained from DATASUS, a database of the national public health system freely available online and maintained by the Department of Informatics of the Brazilian Health System~\cite{datasus}.

\section*{Data Analysis and Results}

In this section we report the results that we obtained by analyzing the empirical data which cover the first 33 years of the AIDS infection in Brazil. Specifically, we investigate the annual absolute frequency $y_{i,t}$, defined as the number of new AIDS cases diagnosed in the year $t$ at a given city $i$. It is valid noting that the System Data Registry of DATASUS hosts only mandatory cases of AIDS reported by health professionals to health services. Therefore, the number of HIV positive patients was not available in DATASUS until 2012, but only the number of AIDS cases. Altogether, we analyze data ranging from 1980 up to 2012 of 5138 Brazilian cities with at least one AIDS case in the aforementioned period. Up to the end of 2012, the infection had already reached more than 5000 cities ($\sim 90\%$ of the Brazilian cities). We start with an overview of the spatial spreading of the virus. Next, we focus on the temporal evolution of the infection in Brazilian cities.

\subsection*{Profile of AIDS in Brazil}

\begin{figure}[!ht]
\begin{center}
\includegraphics[width=.7\textwidth]{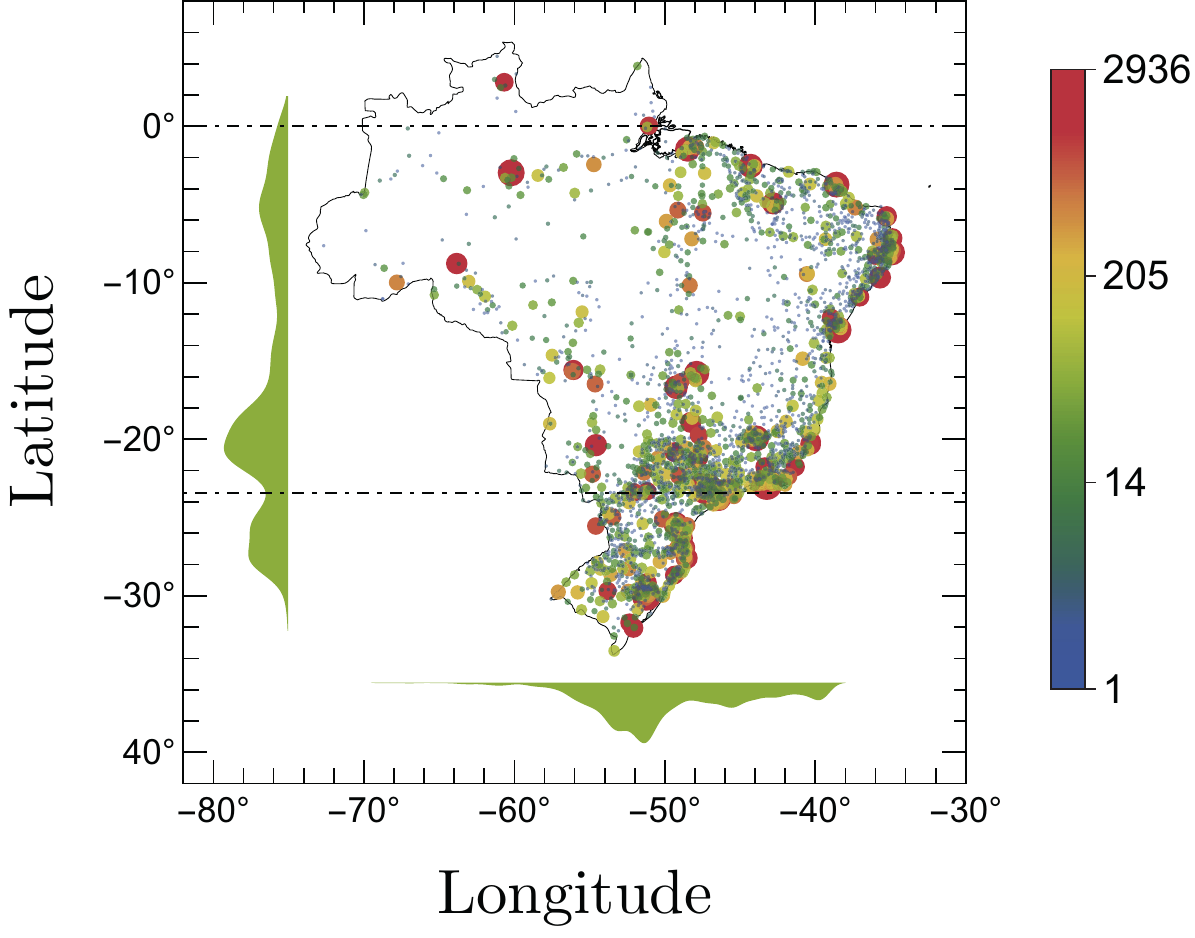}
\end{center}
\caption{
{\bf Figure 1. New cases of AIDS in Brazil reported in 2012.} 
The radius of the circles are proportional to the logarithm of the absolute frequency $y_{i,t}$. 
The histograms along the axis represent the dependence of the distribution of the AIDS cases on the geographic position (latitude and longitude) of the cities. 
Besides reflecting in a great extent the population distribution, this figure also provides a general information concerning the spatial spreading of the epidemics over the country. 
}
\label{growth}
\end{figure}

The first AIDS case in Brazil was identified in the year 1980 in S\~{a}o Paulo, the most populous Brazilian city (southern region). 
In the next few years, new satellite cases came out and the infection started spreading to a larger area. 
In this mean time, the infection spread to other cities and the whole country was already taken by the end of the 1990s. 
The current status of the AIDS infection in Brazil is depicted in Figure~\ref{growth}, where the size of the circles is proportional to the logarithm of $y_{i,t}$ and the histograms along the axis detail the annual absolute frequencies as function of the geographic position of the cities. 
A chronological evolution of the AIDS cases per year among Brazilian cities is available as a supplementary figure (Figure~\ref{chronology}). 
As clearly illustrated in Figure~\ref{growth}, the infection is concentrated in some key cities located mainly next the Atlantic littoral, where is concentrated most of the Brazilian population. 

\subsection*{Time evolution of $y_{i,t}$ total}

\begin{figure}[!ht]
\begin{center}
\includegraphics[width=\textwidth]{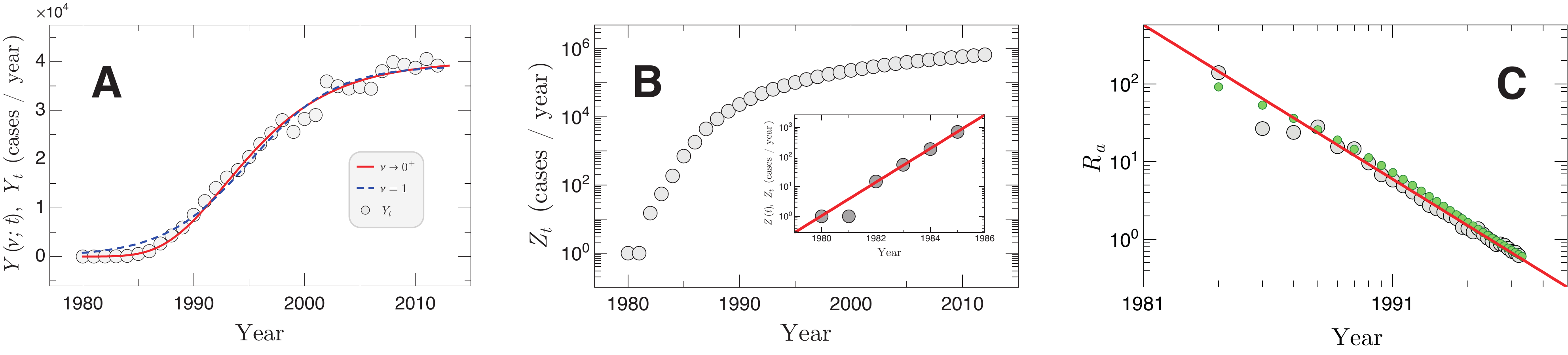}
\end{center}
\caption{
{\bf Figure 2. Time evolution of the epidemics.} 
(a) Growth of $Y_{t}$ within the period 1980--2012 for 5138 Brazilian cities. 
The dashed blue line is Equation~(1) with the parameters $a = 756$, $b = 51.8$, $\beta = 0.26$ and $\nu = 1$. 
The continuous red line is Equation~(1) with the parameters $a = 1$, $b = 40132$, $\tilde{\beta} = 0.18$ and $\nu\rightarrow 0^{+}$. 
(b) Growth of the corresponding accumulated number of cases $Z_{t}$ shown in log-lin scale. 
The inset represents the first points of the curve of $Z_{t}$, corresponding to the period 1980--1985. 
The solid line is a fit in this region, given by $Z(t) \propto e^{r\, t}$ with $r = 1.31$. 
(c) Decay of the actual reproduction number $R_{a}$ for the whole period shown at log-log scale. The solid line is a linear fit (in log-log scale) giving a power law exponent $\lambda = 1.99$. 
The green points correspond to the numerical solution of the continuous equation $R_{a} (t) = Y(t)/\int_0^{t-1} d\tau\, Y(\tau)$ with $Y(t)$ for the Gompertz model, connecting the fit of Equation~(1) to the patterns of the data. 
}
\label{dynamics}
\end{figure}

First we considered $Y_{t}$, the total annual absolute frequency at each year $t$, obtained by summing $y_{i,t}$ over the total of $N$ Brazilian cities, $Y_{t}=\sum_{i=1}^{N} y_{i,t}$. 
Figure~\ref{dynamics}A shows the temporal evolution of $Y_{t}$ and compares it with the curve
\begin{equation}
Y(\nu;\, t)=\frac{b\, a}{\left[1 + (b^{\nu}-1)\, e^{-\nu\beta\, t}\right]^{\frac{1}{\nu}}}\,,
\label{logistic}
\end{equation}
for different non-negative values of $\nu$.\footnote{As a distinction from the curves continuous in time, when we are referring to the data, we have always used a subscript $t$, emphasizing their discreteness nature. }
This curve is obtained as the solution of the Richard's differential equation and represents a generalized logistic growth~\cite{nelder1961,gaillard1997}.  
Classically, logistic-like equations have a vast application in systems related to the dynamics of populations where there is a strong competition of increasing and establishing factors, exactly as in the case of diseases growing and spreading~\cite{murray}. 
A dimensional analysis reveals that $a$ has the dimension of $cases/year$, $\beta$ goes with $years^{-1}$, while $b$ and $\nu$ are dimensionless. 
The particular case $\nu=1$ corresponds to the usual logistic curve. 
For this case, a least-squared regression to the data\footnote{For all the fits through out this work we have set 1980 as $t=0$. } 
(see the dashed curve on Figure~\ref{dynamics}A) leads to $a = 756\pm 490$, $b = 51.8\pm 32$ and $\beta = 0.26\pm 0.05$ ($R^2 = 0.996$), where the uncertainties correspond to 99\% confidence intervals. 
This type of curves have also been identified in SI (susceptible-infected) simulations of the initial evolution of epidemic outbreaks in growing scale-free networks with local structures~\cite{ni2008}. 
Similar curves has also been identified in networks whose nodes are not only the media to spread the virus, but also to disseminate their opinions about it~\cite{ni2011}. 
In the AIDS spreading case, the establishment of an almost constant growth per year after 1999 (Figure~\ref{dynamics}A) could be associated with factors such as the impact on using ARV therapies as well as the extensive broadcast prevention campaigns realized in Brazil~\cite{grangeiro2010}. 
It is also evident that the limit $\nu \rightarrow 0^{+}$ leads to the Gompertz model\footnote{For reason of convergence, one should have $\tilde{\beta} = \frac{\beta}{\nu}$.}, that was originally proposed to model mortality of an aging population~\cite{gompertz1825}. Almost century later, this approach still has applications in biological growth curve~\cite{winsor1932}, in survival analysis~\cite{coe2005} and in regeneration~\cite{wallenstein2004}.
A least-squared regression of the Gompertz curve with $y_{0} = 1$ to the data (see the continuous curve on Figure~\ref{dynamics}A) leads to $b = 40132\pm 1689$ and $\tilde{\beta} = 0.18\pm 0.01$ ($R^2 = 0.997$), where the uncertainties correspond to 99\% confidence intervals. 
It is noticeable that the Gompertz model is better than the generalized logistic model (for different values of $\nu$) when describing the beginning of the curve, but looking at the residuals of fit, both models are in good agreement with the data.

Figure~\ref{dynamics}B shows $Z_{t}$, the correspondent accumulated annual absolute frequency defined mathematically as $Z_{t}=\sum_{j=1}^{t}Y_j$. 
In a few words, $Z_{t}$ represents the total number of AIDS cases diagnosed in Brazil until the year $t$. 
For small times, $Z_{t}$ exhibits an exponential growth, $Z(t) \propto e^{r\, t}$ (inset in Figure~\ref{dynamics}B), where $r$ has the unity $years^{-1}$. 
The limit $Z(t)\sim e^{r\,t}$ can be recovered (analytically) from Equation~(1) when $\beta\, t$ is very small. 
A numerical comparison with the data reveals a relative error of 0.8 (for the logistic model) or of 0.5 (for the Gompertz model) for small times. 
In this background, $r$ can be considered an estimate of the intrinsic growth rate 
describing the initial growth of the epidemic, which underlies the entire spreading of the infection in the surrounding regions. 
Naturally, the intrinsic growth rate $r$ can be estimated from this approach since in the beginning of the epidemic $Z_{t}$ practically corresponds to the prevalence of AIDS infection cases. 
A linear regression to the data leads to $r = 1.31\pm 0.08$~$years^{-1}$ (99\% confidence interval; $R^{2} = 0.999$) for the AIDS epidemic among Brazilian cities. 
Similar analysis for AIDS epidemic in France, in the UK, and in Western Germany have led to $r \simeq 1.15$, $r \simeq 1.21$, and $r \simeq 2.15$ respectively~\cite{hiroshi2010}.
As for HIV/AIDS epidemic among homosexual population in England and Wales (1981--1996), it was found $r \simeq 0.9$~\cite{gran2008}.

In contrast to the intrinsic growth rate $r$, reproduction numbers describe a more local or individual quantity. 
The actual reproduction number $R_{a}$ is defined as the average number of secondary cases per case to which the infection was actually transmitted during the infectious period in a population. 
An estimate of $R_{a}$ can be obtained assuming that $R_{a}=(Y/Z)\,D$, where $D$ is the average length of the HIV infectious period ($\simeq 10$ years)~\cite{bailey1988,gran2008}. 
Since the ratio ($Y_t/Z_{t-1}$) can be associated to a transmission index, reproduction numbers carry important indicators like the threshold for pandemics. 
Figure~\ref{dynamics}C shows the temporal evolution of $R_{a}$ in Brazil in comparison with a power law decaying in the form
\begin{equation}
R_{a}(t) \propto t^{-\lambda}, 
\label{ra}
\end{equation}
with $\lambda = 1.99\pm 0.07$ (99\% confidence interval; $R^{2} = 0.995$). 
For sure, Equation~(2) can be associated from Equation~(1) with the assumption that accumulating a discrete time series is a good approach for an integration process over a continuous variable. 
In particular, the numerical solution to the continuous approach $R_{a} (t) = Y(t)/\int_0^{t-1} d\tau\, Y(\tau)$ with $Y(t)$ given by Equation~(1) also leads to consistent results within the confidence intervals (see the green points in Figure~\ref{dynamics}C). 
Using incidence data of HIV/AIDS epidemic only among homosexual population in England and Wales (1981--1996), a power law pattern could not be identified~\cite{gran2008}. 
Clearly, that data are a small fraction of the population and have severe limitations in time and space, which does not happen with the Brazilian data, which have continental proportions, including more than 5000 urban centers and more than 30 years of data.

\subsection*{Distribution of $y_{i,t}$ in the cities}

\begin{figure}[!ht]
\begin{center}
\includegraphics[width=\textwidth]{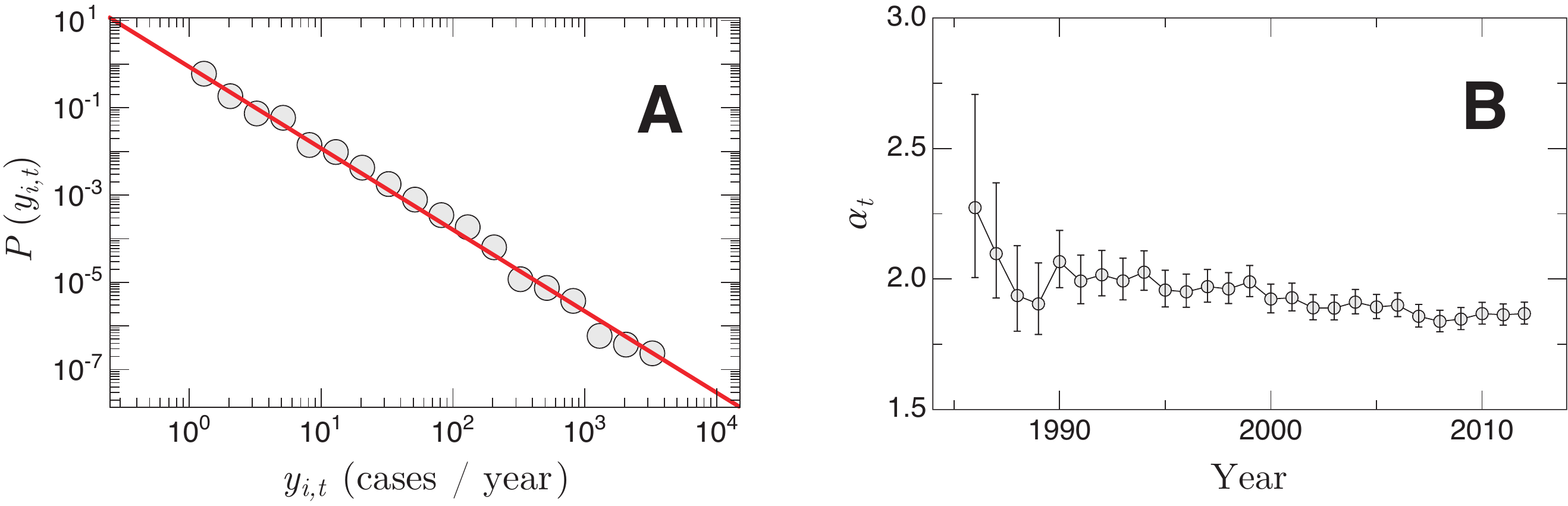}
\end{center}
\caption{
{\bf Figure 3. PDF of the absolute frequency among cities. }
(a) Log-log plot of the probability density function, $P(y_{i,t})$, where $y_{i,t}$ corresponds to the number of AIDS cases diagnosed in 2012 on 2891 Brazilian cities. The solid line has slope 1.87 and was obtained by a maximum likelihood fit to the data. 
(b) Temporal evolution of the power law exponent $\alpha_t$ for the period 1986--2012. The error bars correspond to 99\% bootstrapped confidence intervals. 
}
\label{powerlaw}
\end{figure}

As usually happens in a epidemic spreading, the AIDS infection had focus in a specific region and then spread to a larger area (see Figure~\ref{chronology}). 
During the spreading process, individuals in many different urban centers get infected and the radius of the infection grew fast. 
A natural measure characterizing such process is a function indicating the number of cities with the same number of AIDS hosts. 
We investigated that through the probability density function (PDF) of the annual absolute frequency of AIDS among Brazilian cities $P(y_{i,t})$ for fixed years. 
Figure~\ref{powerlaw}A shows $P(y_{i,t})$ of the set of all the $y_{i,t}$ in the year 2012 in comparison with the power law decay 
\begin{equation}
P(y_{i,t}) \propto y_{i,t}^{-\alpha_t}. 
\label{distribution}
\end{equation}
This result indicates that the distribution of the set of $y_{i,t}$ over all the Brazilian cities has a robust long-tailed behavior. 
Similar power laws can also be identified for all the $y_{i,t}$ with $t$ fixed. 
In particular, the Cram\'{e}r-von Mises test assures that for $y_{i,t} > 10$ the null hypothesis that the yearly data is distributed accordingly to power law curves can not be reject at a confidence level of 99\% (the $p$-values lies in the interval $(0.08, 0.88]$). 
The annual evolution of the power law exponent $\alpha_t$ with 99\% bootstrapped confidence intervals is shown in Figure~\ref{powerlaw}B. 
In particular, for the year 2012, $\alpha_{t}=1.87\pm 0.04$ (99\% bootstrapped confidence interval). 
The values for $\alpha_t$ were obtained by a maximum-likelihood estimation. 
In spite of large fluctuations for initial times due to a small quantity of infected individuals, $\alpha_t$ stays practically constant over the whole period ($\alpha_t \simeq 1.9$), but exhibiting a slight trend to decrease if the error bars were disregarded. 

\subsection*{$y_{i,t}$ \textit{vs.} population of cities}

\begin{figure}[!ht]
\begin{center}
\includegraphics[width=\textwidth]{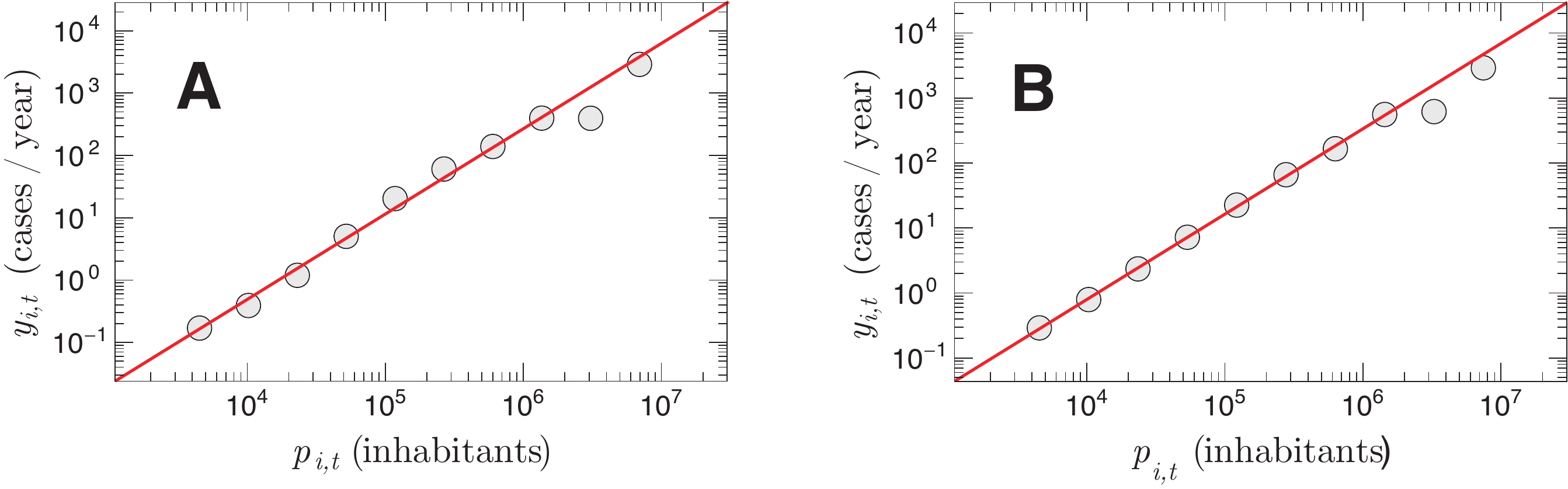}
\end{center}
\caption{
{\bf Figure 4. Allometric relationship between the epidemics and population of the cities. }
$y_{i,t}$ as a function of the population of the city ($p_{i,t}$) shown at log-log scale for fixed years: 
(a) 2000; and (b) 2010. 
The solid lines are linear fits performed for cities with 3000 or more inhabitants. 
The solid lines have slopes (a) 1.37 and (b) 1.31. 
}
\label{allometry}
\end{figure}

The actual number of infected individuals per capita depends on many different factors and can not be expected to be a constant. 
Factors related to high-risk behaviors are connected with the majority of new infection cases. 
Naturally, the frequency of some infection in a given city should be related somehow to the size of that city (as its number of inhabitants). 
To investigate this possible size dependence, we crossed the AIDS data with data on the population of cities obtained from the Brazilian Institute of Geography and Statistics (IBGE)~---~a federal government's institution which performs a national census about every ten years~\cite{IBGE1,IBGE2}. 
Figure~\ref{allometry} shows $y_{i,t}$ of each city as a function of its number of inhabitants $p_{i,t}$. 
In order to obtain a better approach, we selected only the cities with $p_{i,t} \geq 3000$ for two fixed years and considered windows-averaged values. 
Figure~\ref{allometry}A shows both $y_{i,t}$ and $p_{i,t}$ in year 2000, while Figure~\ref{allometry}B shows the analogous plot for the year 2010. 
The curves are well described by the scaling law 
\begin{equation}
y_{i,t} \propto p_{i,t}^{\delta_t}. 
\label{allometryequation}
\end{equation}
In particular, we found $\delta_t = 1.37\pm 0.13$ (99\% confidence interval; $R^{2} = 0.995$) for the year 2000 and $\delta_t = 1.31\pm 0.03$ (99\% confidence interval; $R^{2} = 0.999$) for the year 2010. 
Note that the annual absolute frequency of AIDS cases depends on the population with a non-trivial rule, since $\delta > 1$. 
This super-linear dependence indicates that the infection grows in average faster than the population, meaning that the more populous the city, the greater the density of the infection per capita. 
Similar allometric scaling laws have been identified in many biological systems. 
Classical examples are the metabolic rates scaling with the mass of animals and plants proportions~\cite{west1999,niklas2004}. 
Consistent results were also found for data on new AIDS cases from U.S. related to the period 2002--2003 ($\delta\simeq 1.23$), as well as some social indicators~\cite{bettencourt2007}. 

It is clear that an allometric relationship between two variables is usually outlined by the probability distribution of both the variables. 
Studies in the literature investigate such type of connection. As an example, there is an evident relationship between homicides and population/urban indicators~\cite{alvesplosone2013}. 
In particular, by a direct transformation of probabilities~\cite{bettencourt2012}, it is possible to show that an allometric exponent $\delta$ can be connected with two other power law exponents $\alpha$ and $\beta$ by the equation $\delta =(\beta -1)/(\alpha -1)$. 
In our scaling laws, $\alpha$ represents the exponents obtained by the probability distribution of AIDS cases on the Brazilian cities for each year (see Equation~(3) and Figure~\ref{powerlaw}B) and $\beta$ came from the probability distribution of sizes of the Brazilian cities. 
$\beta \simeq 2.3$ refers to the Zipf law of the Brazilian cities~\cite{malacarne2001,moura2006}. 
Naturally, concerning our data, the exponent $\beta$ can be slightly different because we are considering only the cities that had at least one AIDS case in the corresponding year.

\subsection*{Dynamics of $y_{i,t}$}

\begin{figure}[!ht]
\begin{center}
\includegraphics[width=\textwidth]{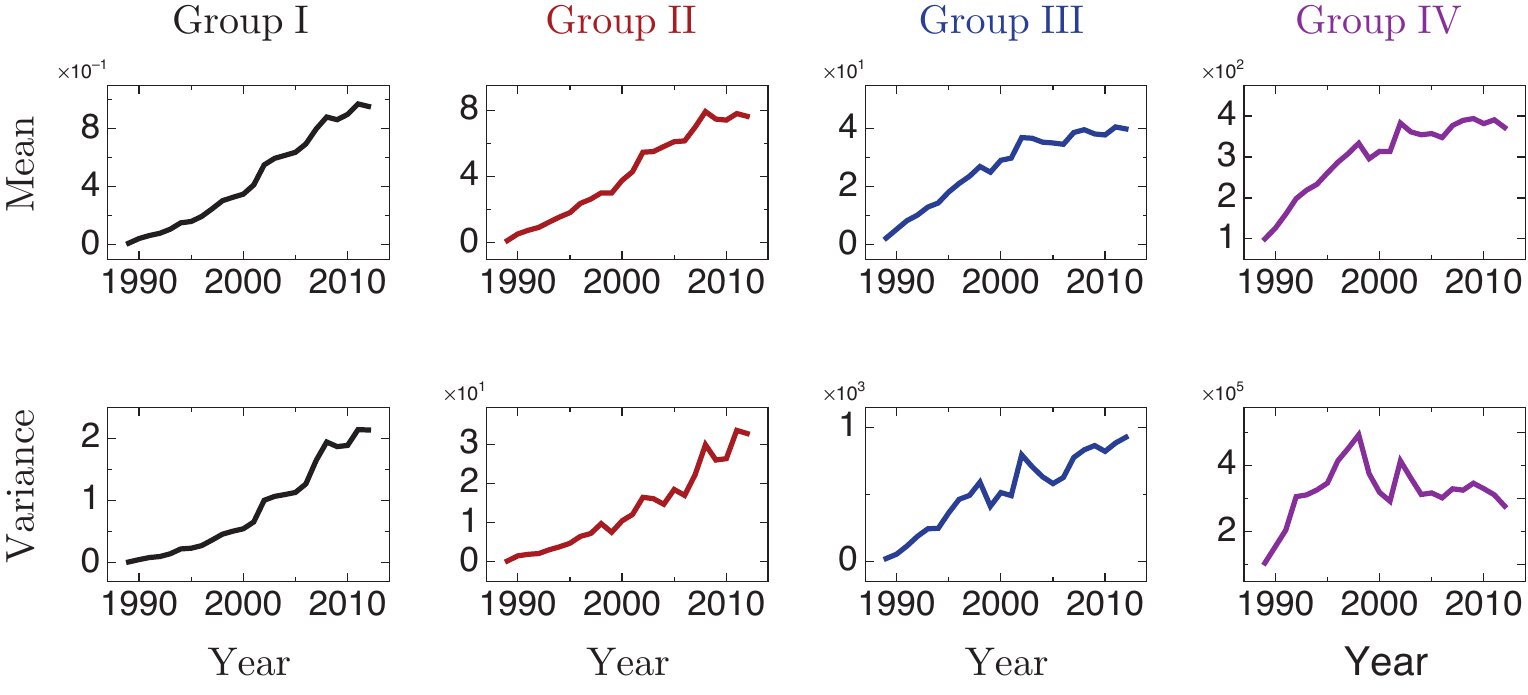}
\end{center}
\caption{
{\bf Figure 5. Moments of the $y_{i,t}$ in Brazilian cities in the period 1989--2012. }
The total of 5138 cities are split in 4 groups:
$\langle y_{i,t}\rangle < 2$ (Group I), 
$2 \leq \langle y_{i,t}\rangle \leq 10$ (Group II),
$10 < \langle y_{i,t}\rangle \leq 84$ (Group III), and
$\langle y_{i,t}\rangle > 84$ (Group IV). 
}
\label{momentos}
\end{figure}

As pointed out before, Brazil is a forefront reference in providing AIDS prevention campaigns and treatment services~\cite{nunn2007}. 
The country has been investing systematically for years. 
For example, only in 2008, Brazilian government invested more than US\$~600 million (around 1/3 of the amount spent in Caribbean and all the three Americas) in HIV domestic and international spending~\cite{UNAIDS}. 
As a consequence, these efforts should have had an impact over the epidemic of AIDS in Brazil. 
To quantify such impact, disregarding other possible contributions, we split all the Brazilian cities with at least one AIDS case since 1989 into 4 groups accordingly to their average absolute frequency per year, denoted by $\langle y_{i,t}\rangle$ in $years$ and defined as an arithmetic mean of $y_{i,t}$: 
Group I accommodates 4029 cities with $\langle y_{i,t}\rangle < 2$. 
Group II encompasses 565 cities with $2 \leq \langle y_{i,t}\rangle \leq 10$. 
Group III accommodates 259 cities with $10 < \langle y_{i,t}\rangle \leq 84$, and, finally, Group IV covers 48 cities with $\langle y_{i,t}\rangle > 84$. 
It is evident from Figures~\ref{allometry} that, $y_{i,t}$ has a well-defined allometry with the population of cities. 
For this reason, separate groups of cities accordingly to $\langle y_{i,t}\rangle$ is also equivalent to separate $y_{i,t}$ by population-sizes, apart from random fluctuations. 
In Figure~\ref{momentos} we show average values and variance for each one of these groups. 
We can see that for small cities (Groups I and II), the average number of cases and the respective fluctuations are still growing, while large cities (Groups III and IV) exhibit a different behavior: the number of cases are already decreasing, as well as the fluctuations. 
This could reflect that the intensive broadcast programs of the Brazilian government against the infection has been giving results mainly in the largest urban centers, where is concentrated most of the cases. 
Intense treatment therapies started in Brazil at the beginning of the 1990s and had a peak in 1996 with the introduction of HAART (Highly Active Antiretroviral Therapy) which substantially reduces the infectiousness of people living with AIDS~\cite{porco2004,fonseca2007}. 
Estimates suggest that more than 1.2~million of life-years are estimated to have been gained in Brazil between 1996 and 2009~\cite{UNAIDS}. 
Their impact over the frequency of new infection cases is evident in cities of the groups III and IV since the end of the 1990s. 
Many therapeutic approaches have been under constant investigation in order to develop vaccines against the HIV virus or at least control some effects of the infection. 
But ongoing research is unavoidable to identify a solution to this devastating worldwide epidemic. 
\begin{figure*}[h]
\end{figure*}

\section*{Discussion and Conclusions}

In the present work, we showed that the AIDS epidemic in Brazil (1980--2012) exhibits the following growth patterns:
\textit{(i)} logistic-type growth of total AIDS cases per year with an exponential regime in the first few years of the AIDS spreading and \textit{(ii)} power law-type decaying of the actual reproduction number; \textit{(iii)} power law behavior with a robust exponent in the PDF of the annual absolute frequency among cities; \textit{(iv)} a robust allometric relationship between the epidemic and the population of the cities; and, finally, \textit{(v)} different profile for the temporal evolution of the epidemics depending on the average absolute frequency of cases in the cities. 

The result shown in Figure~\ref{dynamics}A suggests a logistic-like growth of the total AIDS cases per year in Brazil. 
Typically such curves are associated with a competition of increasing and establishing agents. 
For this reason, they have a vast application in the dynamics of populations, including the spreading of infectious diseases. 
In the AIDS case, the establishing agents seems to be connected to the impact of the large-scale control and programs of HIV/AIDS intervention in Brazil, like the extensive broadcast prevention campaigns and the evolution with the antiretroviral therapies~\cite{fonseca2007,grangeiro2011,nunn2007}. 
Looking carefully at Figure~\ref{dynamics}A it is possible to identify that, mainly after the year 1999, there is a tendency to saturation of $Y(t)$. 
Such region represents the period where the impact of the efforts against the AIDS epidemics became more intense. 

It is well-known that the number of cases of a new infection in a confined area grows exponentially. 
For the AIDS epidemics in Brazil, we found that this exponential provides the parameter $r=1.31$ (inset of Figure~\ref{dynamics}B) which are in good agreement with the growth of AIDS in some Europeans countries~\cite{hiroshi2010}. 
For small time scales (typically smaller than the characteristic time for infection-related deaths) the prevalence of an infection can be associated to the whole number of infected individuals. 
Using this assumption, we obtained Figure~\ref{dynamics}C, which shows that the growth in the absolute frequency of cases is decreasing with time. 
This result is consistent with the literature~\cite{UNAIDS2} and predicts that the infection is growing slower with time. 
It should be noted that the fluctuations after \textit{ca.} 1992 occurs because it is the region where the small time scale is already broken (\textit{ca.} 10 years for AIDS). 

We also showed that the distribution of the AIDS epidemics in small spatial scales (over cities) follows a power law behavior for the whole period with an decreasing exponent typically around 1.9 and systematically below the inverse-square law ($P(y)\propto y^{-2}$) widespread in the nature. 
Yet, the infection over cities scales with the population in that region. 
We showed through Figure~\ref{allometry} that the scale rule follows a super-linear allometric pattern ($\delta \simeq 4/3$). 
In contrast, mortality rates from influenza and pneumonia in US cities surrounding the 1918 pandemic scales linearly with the population size ($\delta \simeq 1$)~\cite{acunasoto2011}. 

In a general scenario, new AIDS cases in Brazil are running to a plateau (Figure~\ref{dynamics}A). 
Assuming that this behavior is mainly due to the results of the programs and strategies, against the AIDS infection, we focused on the level of cities, elaborating Figure~\ref{momentos}. 
By using it, we showed that the first results of the efforts against the infection occurred mainly in urban centers with an average of more than 10 AIDS cases per year (Groups III and IV). 
That is, the infection is not expected to grow uncontrollably in such cities; in particular, it is expected to lose its strength in the largest urban centers (average of more than 84 AIDS cases per year). 
Naturally, this result is reinforced by the power law decaying in the actual reproduction number (Figure~\ref{dynamics}C), what indicates that the fraction of number of new cases per case is decreasing and whose effect is noted mainly in the larger cities because the fluctuations of cases appears to be smaller. 
On the other hand, the average number of new infections per year (as well as their fluctuations) in cities of the groups I and II are still growing. 
The data is not long enough in time to identify if this fact is just a delay period or if such cities have not experienced sufficiently the Brazilian control strategies against AIDS. 
In any case, they merit further attention from the Brazilian health authorities. 
In a general summary, we proposed in this work an alternative approach to the usual methods of epidemics analysis, emphasizing how the dynamics of the virus at the level of municipalities behaves collectively.

\section*{Acknowledgments}

The authors thank for the financial support of the Brazilian agencies CNPq, CAPES, and the National Institute of Science and Technology for Complex Systems. 
F.J.A. is especially grateful to Capes/Funda\c{c}\~{a}o Arauc\'{a}ria for the grant number 87/2014.


\renewcommand\thefigure{S\arabic{figure}}
\setcounter{figure}{0}

\newpage

\section*{Supplementary Figure}

\begin{figure}[!ht]
\begin{center}
\includegraphics[width=.875\textwidth]{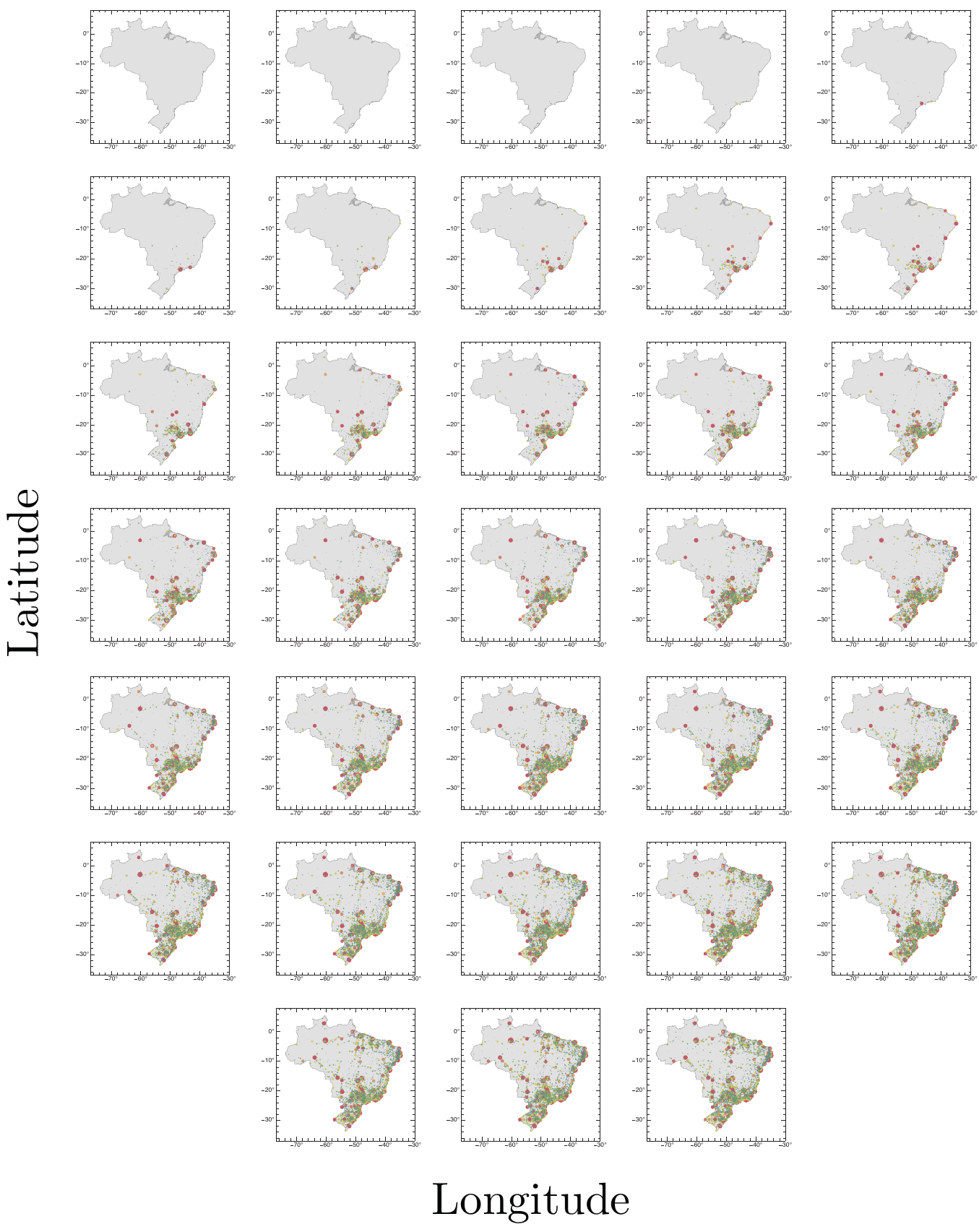}
\end{center}
\caption{
{\bf Figure S1. Chronological evolution of the AIDS epidemics among Brazilian cities. }
Besides reflecting in a great extent the population distribution, this  animation also provides a general information concerning the spatial spreading of the epidemics over the country. 
}
\label{chronology}
\end{figure}

\end{document}